\documentstyle[12pt]{article}
\begin{document}

\newcommand{\ksi}{\xi}

\def\figcap{\section*{Figure Captions\markboth
        {FIGURECAPTIONS}{FIGURECAPTIONS}}\list
        {Figure \arabic{enumi}:\hfill}{\settowidth\labelwidth{Figure
999:}
        \leftmargin\labelwidth
        \advance\leftmargin\labelsep\usecounter{enumi}}}
\let\endfigcap\endlist \relax

\begin{titlepage}

\begin{center}
{\bf Budker Institute of Nuclear Physics}
\end{center}

\vspace{1cm}

\begin{flushright}
BINP 93-115\\
December 1993
\end{flushright}

\bigskip

\begin{center}
{\bf NUCLEAR ANAPOLE MOMENTS}
\end{center}
\begin{center}
{\bf IN SINGLE-PARTICLE APPROXIMATION}
\end{center}

\begin{center}
V.F. Dmitriev\footnote{e-mail address: dmitriev@inp.nsk.su},
I.B. Khriplovich\footnote{e-mail address: khriplovich@inp.nsk.su}
and V.B. Telitsin\footnote{e-mail address: telitsin@inp.nsk.su}
\end{center}
\begin{center}
Budker Institute of Nuclear Physics, 630090 Novosibirsk,
Russia
\end{center}

\bigskip

\begin{abstract}
Nuclear anapole moments of $\;^{133}$Cs, $\;^{203,205}$Tl, $\;^{207}$Pb,
$\;^{209}$Bi are treated in the single-particle approximation. Analytical
results are obtained for the oscillator potential without spin-orbit
interaction. Then the anapole moments are calculated numerically in a
Woods-Saxon potential which includes spin-orbit interaction. The results
obtained demonstrate a remarkable stability of nuclear anapole moment
calculations in the single-particle approximation.
\end{abstract}

\vspace{7cm}

\end{titlepage}

\section{Introduction}

The existence of parity nonconservation (PNC) in atoms is firmly
established at present (see, e.g., Ref. \cite{khr}). To be precise, only
the nuclear-spin-independent PNC effects in heavy atoms have been observed
up to now. Just these effects are enhanced as $Z^2 Q$. The last enhancement
factor, the so-called weak nuclear charge $Q$ which is numerically close to
the neutron number $N=A-Z$, is due to the fact that in the
nuclear-spin-independent phenomena all the nucleons act coherently.

As to the atomic PNC effects dependent on nuclear spin, they evidently lack
this coherent enhancement and are therefore much smaller. There are strong
reasons to expect that these effects are dominated by contact
electromagnetic interaction of electrons with nuclear anapole moment (AM)
\cite{fk,fks}.

Anapole is a new electromagnetic moment arising in a system without centre
of inversion \cite{zel}. It exists even in such a common object as a chiral
molecule in a state with nonvanishing angular momentum \cite{kp}. Nuclear
anapole moment is induced by PNC nuclear forces.

The electromagnetic PNC interaction of electrons with nuclear AM is
conveniently characterized in the units of the Fermi weak interaction constant
$G=1.027\times 10^{-5} m^{-2}$ \footnote{The system of units with
$\hbar=1, c=1$ is used; $m$ is the proton mass.} by a dimensionless
parameter $\kappa$.  (Its definition is given in the next section.) A
closed analytical expression has been obtained for this constant \cite{fks}
within the nuclear shell model under some extra simplifying assumptions.
In particular, for $^{133}$Cs this model prediction is (at the numerical
value of the effective PNC nuclear constant accepted in the paper quoted)
\begin{equation}
\kappa(^{133}\mbox{Cs})=0.33.
\end{equation}
More reliable numerical calculations with the Woods-Saxon potential
including the spin-orbit interaction result at the same PNC nuclear
constant in \cite{fks}
\begin{equation}\label{eq:ws}
\kappa(^{133}\mbox{Cs})=0.25.
\end{equation}

The constant $\kappa(^{133}\mbox{Cs})$ has been calculated also in
Ref. \cite{hhm}. In that paper only the $\pi$-meson-exchange contribution
to the P-odd nuclear forces has been included. This contribution
constitutes roughly a half of the strength of these forces according to
the estimates accepted in Ref. \cite{fks} (the second half is due mainly
to the P-odd $\rho$-exchange). No wonder therefore that the number $0.14$
obtained in Ref. \cite{hhm} for $\kappa(^{133}\mbox{Cs})$ constitutes
roughly a half of the total result (\ref{eq:ws}). (In Table \ref{tab:k},
containing the summary of theoretical results, in the corresponding entry
we indicate in brackets what to our guess would be the result of Ref.
\cite{hhm} if the $\rho$-exchange were included.)

Later numerical calculations with the oscillator potential \cite{bp} led at
the same value of the P-odd nuclear constant $g_p$ to
\begin{equation}
\kappa(^{133}\mbox{Cs})=0.24.
\end{equation}
Less satisfactory is the agreement between the results for the AM of
$\;^{209}$Bi obtained in Refs. \cite{fks,bp}. As to the AM
of $\;^{203,205}$Tl the disagreement between the predictions made in
Refs. \cite{fks,bp1} is even stronger. (The values of
$\kappa(^{203,205}\mbox{Tl}), \;\;\kappa(^{209}\mbox{Bi})$ at the same
magnitude of $g_p$
as accepted in Ref. \cite{fks} is not explicitly given in
Refs. \cite{bp,bp1}. So, in the corresponding entries of Table \ref{tab:k} we
present in brackets our extrapolation of
$\;\kappa(^{203,205}\mbox{Tl}),\;\kappa(^{209}\mbox{Bi})$,
obtained in Refs. \cite{bp,bp1}, from their values of $g_p$ to that
accepted in Ref. \cite{fks}.)

Experiments aimed at the detection of nuclear AM in cesium, thallium, lead,
bismuth are underway in many groups.
The first evidence of a nuclear-spin-dependent P-odd effect has been seen
already in cesium \cite{wie}. The result of this experiment is
\begin{equation}
\kappa(^{133}\mbox{Cs})=0.72(39).
\end{equation}
Therefore, detailed theoretical investigation of various contributions to
nuclear AM looks quite relevant.

The present paper is organized as follows. In the next section we describe the
constant-core-density approximation which leads to a closed analytical
expression for nuclear AM, and introduce some necessary notions and
relations.  Then we consider somewhat more sophisticated model which still
allows for an analytical treatment. This model gives some idea about the
corrections to the above leading approximation. But the most essential part
of the paper is
devoted to serious numerical calculations using a realistic
description of the core density. The nucleon wave functions and Green's
functions are obtained with a Woods-Saxon potential which includes the
spin-orbit interaction. The contribution of the current generated by the
spin-orbit interaction (omitted in Ref. \cite{fks}) is taken into account.
As distinct from Refs. \cite{bp,bp1}, the spin-orbit interaction itself is
treated beyond the perturbation theory.

Though the deviations of the contact and
spin-orbit currents from naive potential expressions are included,
but in all other respects we restrict throughout the present paper to the
single-particle approximation, that of a valence nucleon above a
spherically-symmetrical core.  The many-body effects are certainly of
importance for nuclear anapole moments. We shall consider them in
future publications.

\section{The leading approximation
\newline for nuclear anapole moment}

PNC interaction in a system (atomic nucleus is now of interest to us)
mixes opposite parity states of the same total angular momentum and
creates in it a spin helical structure \cite{khr,zel}. In this way such a
system with nonvanishing magnetic moment acquires a specific configuration
of magnetic field, of the type created by the toroidal winding. This is
what is called anapole \cite{zel}.

The AM vector can be conveniently defined as \cite{khr,fk,fks}
\begin{equation}\label{eq:a}
\vec{a}=-\pi \int d\vec{r} r^2 \vec{j}(\vec{r}\;)
\end{equation}
where  $\vec{j}(\vec{r}\;)$ is the current density operator. The
vector-potential produced by the AM is
\begin{equation}\label{eq:vpg}
\vec{A}(\vec{r}\;)=-[\vec{a}\Delta-\vec{\nabla}(\vec{a}\;\vec{\nabla})]
\frac{1}{4\pi r}\rightarrow \vec{a}\delta(\vec{r}\;).
\end{equation}
We omit in the last expression the term $\vec{\nabla}(\vec{a}\;\vec{\nabla})
1/(4\pi r)$, which can be obviously eliminated by a gauge transformation.

When calculating the AM of a heavy nucleus we restrict ourselves to the
shell model and one-particle approximation (both in what we will call the
leading approximation and beyond it). We shall assume that the nuclear
spin $\vec{I}$ coincides with the total angular momentum of an odd valence
nucleon, while the other nucleons form a core with the zero angular
momentum. The effective P-odd potential for an external nucleon can be
presented as follows:
\begin{equation}\label{eq:w}
W=\frac{G}{\sqrt{2}}\;\frac{g}{2m}\;\vec{\sigma}[\vec{p}\rho(r)
+\rho(r)\vec{p}\;].
\end{equation}
Here $\vec{\sigma}$ and $\vec{p}$ are respectively spin and momentum
operators of the valence nucleon, $\rho(r)$ is the density of nucleons in
the core normalized by the condition $\int d\vec{r}\rho(r)=A$ (the atomic
number is assumed to be large, $A\gg 1$). The numerical value of the
dimensionless constant $g_p$ in the case of an external proton is perhaps
close to $4 - 5$ (see. below). For an external neutron the corresponding
constant $g_n$ is smaller, most probably $g_n\ll 1$.

The leading approximation for the AM of a heavy nucleus corresponds to the
assumption that the density $\rho(r)$ is constant throughout the space and
coincides  with the mean nuclear density $\rho_0$. This approximation,
first used in Ref. \cite{cur}, is reasonable if the wave function of the
external nucleon is mainly localized in the region of the core. The
Schr\"{o}dinger equation for the external nucleon
\begin{equation} \label{schr}
[-\frac{1}{2m}\Delta + U(r) + W(\vec{r}\;)]\psi(\vec{r}\;) = E\psi(\vec{r}\;)
\end{equation}
to first order in $W$ for $\rho(r)=\rho_0=const$ has the elementary
solution
\begin{equation}\label{eq:g}
\psi(\vec{r}\;)=(1-i\frac{G}{\sqrt{2}} g \rho_0 \vec{\sigma}\vec{r}\;)
\psi_0(\vec{r}\;).
\end{equation}
Here $\psi_0(\vec{r}\;)$ is the unperturbed wave function of the external
nucleon.
It might seem that now interaction (\ref{eq:w}), which is
equivalent to the electromagnetic interaction with a constant vector-
potential $\vec{A}=-(G/\sqrt{2}) (g/e) \rho_0 \vec{\sigma}$, should not
result in any physical effects at all. However, the spin part of the
current density
\begin{equation}\label{eq:js}
\vec{j^{s}}(\vec{r}\;)=\frac{e\mu}{2m}\vec{\nabla}\times(\psi^{\dagger}
\vec{\sigma} \psi),
\end{equation}
($\mu$ is the nucleon magnetic moment) does work due to the
noncommutativity of the $\sigma$-matrices, even in this approximation.
Simple calculations using formulas (\ref{eq:a}), (\ref{eq:g}) and
(\ref{eq:js}) yield
\begin{eqnarray}\label{eq:ag}
\vec{a}=\frac{Gg\rho_0}{\sqrt{2}}\;\frac{2\pi e \mu}{m}\langle r^2\rangle
\frac{K \vec{I}}{I(I+1)}, \nonumber \\
K=(I+1/2)(-)^{I+1/2-l}.
\end{eqnarray}
Here $l$ is the orbital angular momentum of the external nucleon. As to
its mean square radius $\langle r^2 \rangle$, it coincides to good
accuracy with the squared charge radius of the nucleus
\begin{equation}\label{eq:msr} r^2_q =(3/5)R^2=(3/5)r^2_0 A^{2/3},
\;\;\;r_0=1.2\; fm.  \end{equation}

It is useful to present also the effective local AM operator which, acting
in the space of nonperturbed wave functions $\psi_0(\vec{r}\;)$, produces the
result (\ref{eq:ag}). This operator is
\begin{equation}\label{eq:lo}
\hat{\vec{a}}=\frac{Gg\rho_0}{\sqrt{2}}\;\frac{2\pi e \mu}{m}\;
[-\vec{\sigma} r^2 + \vec{r}\;(\vec{\sigma} \vec{r}\;)].
\end{equation}

Setting $\rho_0=(4\pi r_0^3/3)^{-1}$, we finally obtain from (\ref{eq:ag})
\begin{equation}\label{eq:al}
\vec{a}=\frac{G}{\sqrt{2}}\frac{9}{10}g\frac{e \mu}{m r_0} A^{2/3}
\frac{K \vec{I}}{I(I+1)}.
\end{equation}
The $A$-dependence of the AM is very natural. Indeed, since the anapole
corresponds to the magnetic field configuration induced by a toroidal
winding, the AM value should be proportional to the magnetic flux, i.e., to
the cross-section area of the torus. This is the origin of $\langle r^2\rangle
$
in formula (\ref{eq:ag}) and of $A^{2/3}$ in (\ref{eq:al}).

Let us turn now to the PNC problem in atoms. The Hamiltonian of the
interaction of an electron with vector-potential (\ref{eq:vpg}) can be
presented as
\begin{equation}
H_a= e\vec{\alpha}\vec{a}\delta(\vec{r}\;)=\frac{G}{\sqrt{2}}
\frac{K\vec{I}\vec{\alpha}}{I(I+1)}\kappa \delta(\vec{r}\;)
\end{equation}
($-e$ is the electron charge, $\vec{\alpha}$ are the Dirac matrices). The
Fermi constant $G$ serves as the natural unit for the AM, that arises in
first order in the weak interaction, and has the dimension cm$^2$. In this
unit a convenient characteristic of the nuclear AM for the atomic PNC
problem is the dimensionless constant $\kappa$. According to (\ref{eq:al})
it equals \cite{fks}
\begin{equation}\label{eq:kl}
\kappa=\frac{9}{10}g\frac{\alpha \mu}{m r_0} A^{2/3}.
\end{equation}
The enhancement $\sim A^{2/3}$ compensates to a large degree the small fine
structure constant $\alpha=1/137$. That is why the nuclear AM is perhaps
the main source of the nuclear-spin-dependent PNC effects in heavy atoms
\cite{fk,fks}. The constants $\kappa$ for nuclei of experimental interest,
as given by simple analytical formula (\ref{eq:kl}), are presented in
Table \ref{tab:k}. Their numerical values for Cs, Tl, Bi correspond to
$g_p=4$, the number assumed in Ref. \cite{fks}. In the next line of Table
\ref{tab:k} the results of numerical calculations \cite{fks} are presented
at the same $g_p$. Those calculations were performed using a realistic
description of the core density $\rho (r)$. The wave function and the
Green's function of the valence nucleon were calculated with a Woods-Saxon
potential which included the spin-orbit interaction. The crude analytical
calculation is obviously in reasonable agreement with that numerical one.

\section{Analytical treatment of nuclear AM in a single particle oscillator
potential.
\newline Contact current contribution}

Our analysis of various contributions to nuclear AM beyond the leading
approximation we will start from a more realistic model which still allows
for an analytical treatment. The model consists in the use of the
oscillator potential for the valence nucleon and in neglect of the
spin-orbit interaction.

Let us consider at first the spin current contribution to the AM.
Substituting expression (\ref{eq:js}) into formula (\ref{eq:a}) and
integrating by parts, we transform the corresponding AM operator to the
following form:
\begin{equation}\label{eq:as}
\hat{\vec{a}_s}=\frac{\pi e \mu}{m}\;\vec{r}\times \vec{\sigma}.
\end{equation}
When using the oscillator potential
\begin{equation}
U(r)=\frac{m\omega^2 r^2}{2}
\end{equation}
for the valence nucleon, its radius-vector transforms to
\begin{equation}\label{eq:co}
\vec{r}=-\frac{i}{m\omega^2}[H,\vec{p}]
\end{equation}
where $H$ is the valence nucleon Hamiltonian. Substituting this expression
into the standard second-order perturbation formula for AM
\begin{equation}
\vec{a}_s=-i\frac{\pi e \mu}{m^2 \omega^2}\sum_n \frac{\langle 0|[H,
\vec{p}\times\vec{\sigma}]|n\rangle \langle n|W|0 \rangle + \langle
0|W|n\rangle \langle n|[H,\vec{p}\times\vec{\sigma}]|0\rangle}{E_0 - E_n},
\end{equation}
we reduce it by means of the completeness relation to
\begin{equation}
\vec{a}_s=-i\frac{\pi e \mu}{m^2 \omega^2} \langle 0|[
\vec{p}\times\vec{\sigma}, W]|0\rangle
\end{equation}
With the explicit form (\ref{eq:w}) for the weak interaction Hamiltonian $W$
this formula can be rewritten after elementary transformations as
\begin{equation}\label{eq:ap}
\vec{a}_s=\frac{Gg}{\sqrt{2}}\,\frac{2\pi e \mu}{m} \frac{K\vec{I}}{I(I+1)}\,
\frac{1}{m^2 \omega^2} \langle 0|\rho p^2 - \frac{1}{2}\rho '\partial_r -
 \frac{l(l+1)}{2Kr}\rho '|0\rangle.
\end{equation}

Let us note here that previously the oscillator-potential model was used
in Ref. \cite{dep} for the investigation of other P-odd nuclear effects.
Then it was applied in Ref. \cite{fks1} to estimate the contribution of core
excitations and to rederive formula (\ref{eq:al})
(which follows indeed from expression (\ref{eq:ap}) at $\rho(r)
=\rho_0=const$ when taking into account the oscillator-potential version
of the virial theorem: $\langle 0|p^2/m|0\rangle = \langle 0|m\omega^2
r^2|0\rangle$).

We pass over now to the orbital contribution to the AM. As it was
demonstrated in Ref. \cite{fk}, the AM of a charged particle in a
spherically symmetrical potential can pe presented in the following form:
\begin{equation}\label{eq:a0}
\vec{a}=-i\frac{2\pi e}{m}(\mu - \frac{1}{3}) \frac{K\vec{I}}{I(I+1)}
\sum_n \eta_{0n} r_{0n} - \frac{2\pi}{3}
\langle 0|r^2 \vec{j}_c^0 - \vec{r}(\vec{r} \vec{j}_c^0)|0\rangle.
\end{equation}
Here $\eta_{0n}$ is the P-odd admixture of an intermediate state
$|n\rangle$ to the initial one $|0\rangle$ (this coefficient is purely
imaginary), $r_{0n}$ is the matrix element of $r$ between those states,
$\vec{j}_c^0$ is the contact current operator:
\begin{equation}\label{eq:cn}
\vec{j}_c^0=ie[W,\vec{r}\;]=\frac{Gg}{\sqrt{2}}\frac{e}{m} \rho \vec{\sigma}.
\end{equation}
Therefore, to take into account within the simple-minded potential
approach the orbital contribution for a proton we should substitute $\mu -
1/3$ for $\mu$ in formula (\ref{eq:ap}) and add to that expression the
following contact term:
\begin{equation}
\frac{Gg}{\sqrt{2}}\frac{2\pi e}{m} \frac{K\vec{I}}{I(I+1)} \frac{1}{3}
\langle 0|r^2 \rho |0\rangle.
\end{equation}
In the case of a valence neutron formula (\ref{eq:ap}) needs obviously no
modification within the same model.

However, in fact the contact current contribution both for a valence
proton and neutron looks differently. Indeed, let us start from the weak
interaction Hamiltonian for the outer nucleon $a (a=p,n)$ constructed from
corresponding two-body operators (see, e.g., book \cite{khr}):
\begin{equation}\label{2b}
W^{(2)}  =\frac{G}{\sqrt 2} \frac{1}{4m}
\sum_{a,b}\left(\{(g_{ab}\vec{\sigma}_a -
g_{ba}\vec{\sigma}_b)\cdot(\vec{p}_a - \vec{p}_b), \delta(\vec{r}_a -
\vec{r}_b)\} + g^{\prime}_{ab}[\vec{\sigma}_a\times\vec{\sigma}_b]
\cdot\vec{\nabla}\delta(\vec{r}_a - \vec{r}_b)\right),
\end{equation}
where the notation $\{\;\;, \;\;\}$ means anticommutator. After averaging
this expression over the core nucleons we obtain formula
(\ref{eq:w}) with
\begin{equation}
g=g_{ap}\frac{Z}{A} + g_{an}\frac{N}{A}.
\end{equation}
The ``best values'' \cite{ddh} of weak coupling constants in the
one-meson-exchange approximation lead to the following numbers for the
constants discussed \cite{fks,dfst,fts}:
\begin{equation}
g_{pp}=g_{nn}=1.5,\;\;\;g_{pn}=6.5,\;\;\;g_{np}=-2.2,
\end{equation}
and correspondingly to
\begin{equation}
g=g_p=4.5
\end{equation}
for the valence proton in Cs, Tl, Bi, and
\begin{equation}
g=g_n\ll 1
\end{equation}
for the valence neutron in $^{207}$Pb. Here the constants $g_{pp}, g_{nn},
g_p$ and $g_n$ are effective ones, they include already the exchange terms
for identical nucleons. These constants include also additional
suppression factors reflecting long-range and exchange nature of the P-odd
one-meson exchange as well as the short-range nucleon-nucleon repulsion.
Their values agree with those obtained in Refs. \cite{mck,lob} and differ
from those given in Ref. \cite{ah} since we include into the $\rho$-meson
exchange the suppression due to the short-range
repulsion.

The electromagnetic interaction can be introduced by changing
$\vec{p} \Rightarrow \vec{p}- e\vec{A}(\vec{r})$. The current density
operator is then a derivative of a Hamiltonian over $\vec{A}(\vec{r})$.
This is equivalent to change one power of a momentum operator
$\vec{p}_a \Rightarrow e_a \delta (\vec{r} - \vec{r}_a)$ in Hamiltonian.
This procedure leads to the following expression for the contact current
density operator:
\begin{equation} \label{ccc}
\hat{\vec{j}}_c= \frac{\imath}{2}\{ \sum_a [W^{(2)}, e_a\vec{r}_a
],\delta(\vec{r} - \vec{r}_a )\} = \sum_a \vec{j}^a _c (\vec{r}) \delta(
\vec{r} - \vec{r_a}),
\end{equation}
where
\begin{equation}
\vec{j}^a _c (\vec{r}) = \frac{G}{\sqrt{2}}\frac{1}{m}\vec{\sigma}_a
\sum_b (e_a - e_b) g_{ab} \delta (\vec{r} -\vec{r}_a).
\end{equation}

For the valence protons $e_a=e$ and this operator reduces to
\begin{equation}\label{eq:p}
\hat{\vec{j}_c^p}=\frac{G}{\sqrt 2} \frac{e}{m}\vec{\sigma} g_{pn} \sum_n
\delta(\vec{r} - \vec{r}_n).
\end{equation}
For the outer neutron ($e_a = 0$) we get
\begin{equation}\label{eq:n}
\hat{\vec{j}_c^n}=-\frac{G}{\sqrt 2}\frac{e}{m}\vec{\sigma} g_{np}\sum_p
\delta(\vec{r} - \vec{r}_p).
\end{equation}
After averaging formulae (\ref{eq:p}), (\ref{eq:n}) over the core nucleons we
get the following effective contact currents for the valence proton and
neutron:
\begin{eqnarray} \label{cc}
\vec{j}_c^p=\frac{G}{\sqrt 2}\frac{e}{m} g_{pn}\frac{N}{A}
\rho(r) \vec{\sigma}= \vec{j}_c^0 + \vec{j}_c^1,\\ \nonumber
\vec{j}_c^1=-\frac{G}{\sqrt 2}\frac{e}{m}g_{pp}\frac{Z}{A}
\rho(r) \vec{\sigma};
\end{eqnarray}
\begin{equation}
\vec{j}_c^n=-\frac{G}{\sqrt 2}\frac{e}{m}g_{np}\frac{Z}{A}
\rho(r) \vec{\sigma}. \label{ccn}
\end{equation}
They differ obviously from the naive ones, $\vec{j}_c^0$ for a valence
proton (see (\ref{eq:cn})) and zero for a valence neutron.
Technically it is convenient to retain previous, naive results for anapole
moments, supplementing them with the following correction terms:
\begin{equation}
\vec{a}^{p1}= -\pi\langle 0|r^2 \vec{j}_c^1|0\rangle,
\end{equation}
\begin{equation}
\vec{a}^{n1}= -\pi\langle 0|r^2 \vec{j}_c^n|0\rangle
\end{equation}
for valence proton and neutron respectively. In this way we get the
following closed expression for the AM of a nucleus with a valence proton:
\begin{eqnarray}
\label{eq:app} \nonumber
\vec{a} & = & \frac{G}{\sqrt{2}}\frac{2\pi e}{m} \frac{K\vec{I}}{I(I+1)}
\langle r^2 \rangle \\ \nonumber
& \cdot & \langle 0|g_p\{\mu_p [2 \rho  -  r^2\rho/\langle r^2
\rangle  -  \frac{1}{2m\omega (N+3/2)}\rho ' (\partial_r +
 \frac{l(l+1)}{Kr})] \\
& - & \frac{1}{3}[2 \rho  -  2 r^2\rho/\langle r^2
\rangle  -  \frac{1}{2m\omega (N+3/2)}\rho ' (\partial_r +
 \frac{l(l+1)}{Kr})]\}\\  \nonumber
& - & g_p^1 (1-\frac{1}{2K}) r^2 \rho|0\rangle;\\ \nonumber
\end{eqnarray}
$$g_p^1  =  g_{pp} Z/A=0.6, \; \langle r^2 \rangle =(N+3/2)/m\omega$$
where $N$ is the oscillator principal quantum number. In the case of a
valence neutron the result is
\begin{eqnarray}\label{eq:apn} \nonumber
\vec{a} & = & \frac{G}{\sqrt{2}}\frac{2\pi e}{m} \frac{K\vec{I}}{I(I+1)}
\langle r^2 \rangle \\   \nonumber
& \cdot & \langle 0|g_n \mu_n [2 \rho - r^2\rho/\langle r^2
\rangle  - \frac{1}{2m\omega (N+3/2)}\rho ' (\partial_r +
 \frac{l(l+1)}{Kr})] \\
& - & g_n^1 (1-\frac{1}{2K}) r^2 \rho|0\rangle;\\ \nonumber
\end{eqnarray}
$$g_n^1  =  g_{np}N/A=-1.3.$$

To get the numerical values of nuclear AMs we assume for the
core density a step-like profile $f(r)$:
\begin{equation} \label{dens}
\rho(r) = \rho_0 f(r) = \rho_0 \theta(R-r).
\end{equation}
As to the mean square radius of the valence nucleon $\langle r^2\rangle $,
it is natural to identify it in the shell model with the nuclear magnetic
mean square radius $\langle r^2_m\rangle $.  The empirical data on the
latter (referring unfortunately to lighter nuclei only) are presented in
Table \ref{tab:r}.  The observation is that in the
$\;^{41}\mbox{Ca}_{20}$, with a valence neutron, $\langle r^2_m\rangle $
is close within the error bars to the value predicted by the oscillator
model. Meanwhile for nuclei with odd proton, $\;^{45}\mbox{Sc}_{21},
\;^{51}\mbox{V}_{23}, \;^{59}\mbox{Co}_{27}$, its value is much better
approximated by the naive formula (\ref{eq:msr}).

For further calculations it is natural to go over to the usual dimensionless
oscillator variable
\cite{bm}
\begin{equation}\label{eq:msro}
x=r^2/\rho^2, \;\; \rho^2=(m\omega)^{-1}=\frac{4}{5} (\frac{2}{3})^{1/3}
r^2_0 A^{1/3}.
\end{equation}
Then the expectation values entering eqs. (\ref{eq:app}), (\ref{eq:apn}) reduce
to
\begin{equation}\label{eq:ev1}
\langle 0|f(r)|0\rangle =
\frac{\Gamma(n+l+3/2)}{n!\Gamma^2(l+3/2)}\int^{X}_0 dx
e^{-x} x^{l+1/2} F^2(-n,l+3/2,x),
\end{equation}
\begin{equation}\label{eq:ev2}
\langle 0|r^2  f  (r)|0\rangle =
\rho^2 \frac{\Gamma(n+l+3/2)}{n!\Gamma^2(l+3/2)}\int^{X}_0 dx
e^{-x} x^{l+3/2} F^2(-n,l+3/2,x),
\end{equation}
\begin{eqnarray} \nonumber
-\frac{1}{2m\omega(N+3/2)}\langle 0| f  ' (r)(\partial_r +
 \frac{l(l+1)}{Kr})|0\rangle =
2\frac{\Gamma(n+l+3/2)}{(N+3/2)n!\Gamma^2(l+3/2)}e^{-X/2}X^{(l+3)/2} \\
\cdot F(-n,l+3/2,X)(\frac{d}{dX}+\frac{l(l+1)}{2KX})e^{-X/2} X^{l/2}
F(-n,l+3/2,X).
\end{eqnarray}
Here $X=R^2/\rho^2=(5/4)(3/2)^{1/3} A^{1/3}=1.43 A^{1/3},\;F(a,b,x)\;$ is
a degenerate hypergeometric function, $\;n=(N-l)/2$.

We are interested in the following nuclei:

\[
\begin{array}{cllllll}
\mbox{Cs}       & I=7/2,    & l=4,    & K= 4,   & N=4, & n=0, & X=7.304;\\
& & & & & & \\
\mbox{Tl}       & I=1/2,    & l=0,    & K=-1,   & N=4, & n=2, & X=8.423;\\
& & & & & & \\
\mbox{Bi}       & I=9/2,    & l=5,    & K= 5,   & N=5, & n=0, & X=8.492;\\
& & & & & &  \\
^{207}\mbox{Pb} & I=1/2,    & l=1,    & K= 1,   & N=5, & n=2, & X=8.464;\\
\end{array}
\]

For Cs and Bi where $n=0$ the integrals entering expectation values
(\ref{eq:ev1}), (\ref{eq:ev2}) are nothing else but well-known (and
tabulated, see, e.g., \cite{pag}) incomplete
gamma-functions, $\Gamma (l+3/2, X), \Gamma (l+5/2, X)$ respectively. In
the case of Tl and Pb those expectation values can be also reduced to
incomplete gamma-functions.

The numerical results for nuclear AMs, obtained in this model, are
presented in Table \ref{tab:k}. For Cs, Tl and Bi, at the same value of
$g_p$, they are about 10\% smaller than those of the leading
approximation. The correction to the spin current contribution constitutes
about -4\% in Cs and Bi, -7\% in Tl. The total
contribution of the convection and contact currents is negative in cesium,
thallium
and bismuth. As to lead, the contact current contribution, though nonvanishing,
is very small numerically.

\section{Nuclear AM in Woods-Saxon potential.
\newline Spin-orbit current contribution}
Though being convenient for analytical treatment, the step-like density
profile and oscillator model for a single
particle potential are however too crude. Moreover, the above analytical
model does not take
into account the spin-orbit interaction which is, as we will see below,
quite essential for numerical results. So, in this section we will
describe a numerical treatment of a much more realistic description of a
nucleus based on the Woods-Saxon potential including spin-orbit
interaction and on a more realistic description of nuclear density.

The profiles of both density and the central part of nuclear potential are
known to be similar and well described by a Fermi-type function
\begin{equation}
f(r) ={ 1 \over{1 + exp({{r-R}\over{a}})}},
\label{fermi}
\end{equation}
So, the total single-particle potential $U(\vec{r})$ has been chosen
in a standard Woods-Saxon form
\begin{equation}
U(\vec{r}) = U_{0}f(r) +
U_{ls}\frac{1}{r} \frac{df(r)}{dr}(\vec{l}\vec{\sigma }) + U_{C}(r),
\label{wspot}
\end{equation}
where $U_{C}(r)$ is the Coulomb potential of
a uniformly charged sphere.  In order to study stability of AM
calculations against variations of the single particle potentials we shall
use several sets of the Woods-Saxon potential parameters
\cite{bm,chp}.

The solution of
eq. (\ref{schr}) can presented in a form similar to (\ref{eq:g})
\begin{equation}
\psi (\vec{r}) = \psi_{0}(\vec{r}) + \delta \psi (\vec{r}) = \left(
R_{0}(r) - \imath \frac{G}{\sqrt{2}}g\,\rho_{0}(\vec{\sigma} \vec{n})\,
\delta R(r) \right) \Omega_{Ilm}(\vec{n}),
\label{psi}
\end{equation}
where $\Omega_{Ilm}(\vec{n}) $ is a spherical spinor, $\vec{n} = \vec{
r}/r$. The correction $\delta \psi$ is of the parity opposite to that of the
initial state $\psi_0$. The radial function  $\delta R(r)$ is normalized
in such a way
that without spin-orbit potential and for constant density it is
(see (\ref{eq:g}))
\begin{equation} \label{dR} \delta R(r) = rR_{0}(r).
\end{equation}
This correction can be expressed via the Green's function
$G_{Il'}(r,r')$ of the unperturbed radial
Schr\"odinger equation for the orbital angular momentum $l'= 2I - l$:
\begin{eqnarray} \label{dr1} \nonumber
\delta R(r) = -\frac{1}{2m}\int \;r'^{2} dr'\; \left(\frac{d}{dr'}
G_{Il'}(r,r') R_{0}(r') - G_{Il'}(r,r')\frac{d}{dr'}R_{0}(r')\right.  \\
\left. -\frac{2K}{r'} G_{Il'}(r,r') R_{0}(r') \right)f(r').
\end{eqnarray}
After presenting the Green's function through two linearly
independent solutions of the radial Schr\"odinger equation we get
\begin{eqnarray} \label{dr2} \nonumber
\delta R(r) = \frac{u^{(1)}_{Il'}(r)}{r}\int_{r}^{\infty} dr'\;
\left(u_{Il'}^{(2)}(r') u_{0}'(r') -  u_{Il'}^{(2)\prime}(r') u_{0}(r')
 +\frac{2K}{r'}u_{Il'}^{(2)}(r') u_{0}(r') \right)f(r') \\
+\frac{u^{(2)}_{Il'}(r)}{r}\int_{0}^{r} dr'\;
\left(u_{Il'}^{(1)}(r') u_{0}'(r')  -  u_{Il'}^{(1)\prime}(r') u_{0}(r')
 +\frac{2K}{r'}u_{Il'}^{(1)}(r') u_{0}(r') \right)f(r'),
\end{eqnarray}
where $u_0(r)=rR_0(r)$; $u^{(1)}(r)$ and $u^{(2)}(r)$ are solutions
regular at the origin and at the infinity respectively. Those last two
solutions are normalized to the unit
Wronskian: $ u^{(1)}u^{(2)\prime} - u^{(1)\prime}u^{(2)} = 1$.

The results of this calculation of the $\delta R$ for $1h_{9/2}$ proton
state in $ ^{209}$Bi together with the leading approximation (\ref{dR}) are
shown in Fig.1. In the same picture we demonstrate how $\delta R$ is
influenced by the
spin-orbit potential, as well as by the deviation of the density from a
constant one.
The real form of $\delta R$ differs considerably from the leading
approximation,
due mainly to the spin-orbit interaction. As to the realistic density profile,
it makes  $\delta R$ slightly smaller as compared to
the constant density approximation.

All contributions to the anapole moment can be expressed via the radial
correction $\delta R$. The spin current term can be
obtained from eq. (\ref{eq:ag}) by substituting
\begin{equation} \label{as}
(\delta R |r|R_{0}) = \int_{0}^{\infty}r^{2}dr \; \delta R(r)rR_{0}(r).
\end{equation}
for $\langle r^{2}\rangle $.
So, this contribution to the dimensionless anapole constant equals
\begin{equation} \label{as1}
\kappa_{s} = 2\pi g\frac{\alpha \mu \rho_{0}}{m}(\delta R|r|R_{0}).
\end{equation}

To obtain the convection current contribution we substitute
corresponding current density operator
\begin{equation} \label{conv}
\hat{\vec{j}}_{conv}(\vec{r}) =-\imath \frac{e}{2m}\sum_{p}
\{\vec{\nabla}_{p}, \delta (\vec{r} - \vec{r}_{p})\}
\end{equation}
into formula (\ref{eq:a}) which gives
\begin{equation}
\vec{a}_{conv}
= -\pi \langle \delta \psi|\{ \frac{\vec{p}}{m},r^2\}|\psi_{0}\rangle
\end{equation}
and
\begin{equation} \label{conv1}
\kappa_{conv} = -\pi g \frac{\alpha \rho_{0}}{mK}(\delta R|r^2(\frac{d}{dr}
+ \frac{K+2}{r})|R_0).
\end{equation}

When calculating the contact current contributions it
is convenient again to split it into the "naive" part arising from a single
particle potential (and vanishing for a valence neutron) and the part which is
due to the current densities (\ref{cc}), (\ref{ccn}).

The contribution of the total contact current (\ref{cc}) to the AM can be
expressed via radial matrix element as
\begin{equation} \label{cont}
\kappa_c = \kappa_c ^0 + \kappa_c ^1 ,\ \
 \kappa^i _c =\pi g^i \frac{ \alpha \rho_0}{m}
 (1 - \frac{1}{2K})\langle r^2 f(r) \rangle, \;\; i=0,\;1
\end{equation}
where $g^0 _p = g_p,\;\; g^0_n =0$, and $g^1 _{p,n}$ were
defined in (\ref{eq:app}), (\ref{eq:apn}).

One more term in AM originates from the momentum dependence of
spin-orbit interaction. As it will be seen below, this is the most
significant correction to the leading approximation. The spin-orbit term
in a single particle potential (\ref{wspot}) for a proton generates
the electromagnetic current density
\begin{equation} \label{jls0}
\hat{\vec{j^0_{ls} }} = \imath e [U(\vec{r}),\vec{r}] =
 e U_{ls} \frac{df(r)}{dr} \vec{\sigma } \times \vec{n}.
\end{equation}
For an outer neutron the corresponding current density vanishes.
The spin-orbit contribution to the proton AM and to $\kappa$ is respectively:
\begin{equation} \label{als0}
\vec{a^0_{ls}} = 2\pi \frac{G}{\sqrt{2}} eg\rho_0 U_{ls} (\delta R|r^2 f'
(\vec{n}(\vec{\sigma}\vec{n}) -\vec{\sigma})|R_0) =
2\pi \frac{G}{\sqrt{2}} eg\rho_0 U_{ls} (\delta R|r^2 f'|R_0)
\frac{K\vec{I}}{I(I+1)},
\end{equation}
\begin{equation}
\kappa_{ls}^0 =
2\pi \alpha  g\rho_0 U_{ls} (\delta R|r^2 f'|R_0) .
\end{equation}

However, as it was the case with the contact current, this is not the complete
result.  The true spin-orbit current must be obtained from two-particle
spin-orbit interaction which can be written as
\begin{equation} \label{eq:so}
U^{(2)}_{ls} = \frac{1}{2} \sum_{ab} U^{ab}_{ls}\, (\vec{p}_a - \vec{p}_b)
\cdot (\vec{\sigma}_a + \vec{\sigma}_b) \times \vec{\nabla} \delta
(\vec{r}_a - \vec{r}_b).
\end{equation}
Averaging it over nuclear core we obtain the spin-orbit part of the
single-particle potential (\ref{wspot}) with
\begin{equation} \label{uls}
U_{ls} = \left(U_{ls}^{ap} \frac{Z}{A} + U_{ls}^{an} \frac{N}{A}
\right)\rho_0 .
\end{equation}
The spin-orbit interaction constants were fitted in Ref. \cite{sad74}:
\begin{equation} \label{uls2}
U^{pp}_{ls} = U^{nn}_{ls} = 36.6\; MeV\cdot fm^5,\ \  U^{pn}_{ls} = U^{np}_{ls}
 = 134.3\; MeV\cdot fm^5.
\end{equation}
A word of caution is proper here. We will understand expression (\ref{eq:so})
as a local limit for an operator of a nucleon-nucleon interaction with the
proper
tensor structure. For the identical nucleons the exchange interaction doubles
the direct one. The values of $U^{pp}_{ls}, U^{nn}_{ls}$ presented in
(\ref{uls2}) are just those doubled, effective constants.

The spin-orbit current density from the interaction (\ref{eq:so}) is
\begin{equation} \label{jls}
\hat{\vec{j_{ls}}} = \imath \sum_a \, [U^{(2)}_{ls},
e_a\vec{r}_a]\, \delta(\vec{r} - \vec{r}_a) = \sum_{ab}  U_{ls}^{ab}\,
e_a\delta (\vec{r} - \vec{r}_a)\;
(\vec{\sigma}_a + \vec{\sigma}_b)\times \vec{\nabla}\delta(\vec{r} -
\vec{r}_b).
\end{equation}
This expression for the two-particle spin-orbit current density was used
earlier in the description of magnetic properties of a nucleus in
\cite{dt}.  Including the spin-orbit current into the theoretical description
of nuclear magnetic moments and magnetic transition amplitudes improves the
agreement with experiment.

In the current density generated by the $pp$ part of operator
(\ref{eq:so}) the exchange and direct contributions cancel each other, i.e.,
the contact spin-orbit $pp$ interaction does not generate a current at all.
Then, averaging expression (\ref{jls}) over the core nucleons and
separating the single particle contribution (\ref{jls0}), as in the case of
the contact current, we get the following effective spin-orbit current for the
valence proton:
\begin{eqnarray} \label{jlsp}
\vec{j}^p_{ls} =
eU^{pn}_{ls}\rho_0\frac{N}{A}\frac{df(r)}{dr}\vec{\sigma}\times \vec{n}=
\vec{j}^0_{ls}+\vec{j}^1_{ls};\\ \nonumber
\vec{j}^1_{ls}=
-eU^{pp}_{ls}\rho_0\frac{Z}{A}\frac{df(r)}{dr}\vec{\sigma}\times\vec{n}.
\end{eqnarray}
The correction $\vec{j}^1_{ls}$ to the "naive" spin-orbit proton current is
relatively small being suppressed by a factor
$$\ksi=-\frac{ZU^{pp}_{ls}}{ZU^{pp}_{ls}+NU^{pn}_{ls}}$$
as compared to the potential part. This factor constitutes -0.16 for cesium
and -0.15 for thallium and bismuth.

It is noteworthy that even for a valence neutron the spin-orbit current does
not vanish. It is generated by the neutron-proton interaction and equals
\begin{equation} \label{jlsn}
\vec j_{ls}^n = -eU_{ls}^{np}\frac{Z}{A}\, \rho(r) \, \vec{\nabla}\times
\left( \psi^{\dagger }(\vec{r})\vec{\sigma}\psi(\vec{r})\right)
\end{equation}
This expression is similar to the spin current density (\ref{eq:js}) and
renormalizes in an obvious way the magnetic moment of an outer neutron.

It is convenient to single out in the sum of
all contributions to the effective constant $\kappa$
\begin{equation}
\kappa = \kappa_s + \kappa^0_{ls} + \kappa^1_{ls} + \kappa_{conv} +
\kappa^0_c + \kappa^1_c = \kappa^0 + \kappa^1_{ls} + \kappa^1_c,
\end{equation}
the "naive" potential one $\kappa^0$. The corrections to it $\kappa^1_{ls}$ and
$\kappa^1_c$ in fact go beyond the single-particle approximation.
The single-particle contribution can be written in a more compact form,
analogous to (\ref{eq:a0}):
\begin{equation}\label{a0}
\vec{a}^0 = (1-\frac{1}{3\mu })\vec{a}_s
 + \frac{2}{3}
\vec{a^0_{ls}}  - \frac{2\pi }{3}\langle 0|r^2\vec{j^0_c} - \vec{r}
(\vec{r} \vec{j^0_c}) |0 \rangle .
\end{equation}
This expression was used to check the accuracy of our numerical calculations
which turned out very high.

To study the stability of AM calculations under the variations of the
single-particle potential we performed them for two sets of the Woods-Saxon
potential parameters (\ref{wspot}). In both sets the radii and
diffuseness parameters, $R$ and $a$, are the same for the profiles of central
and spin-orbit parts of potential.

The first set is \cite{bm}:
\begin{eqnarray} \nonumber \label{bmls}
R = R_{ls} = 1.27A^{1/3}\; fm,\;\;\; a = a_{ls} = 0.67\; fm, \\
U_0 = (-51 \pm 33\,\frac{N-Z}{A})\; MeV,\;\;\; U_{ls} = -0.35\,U_0,
\end{eqnarray}
where the signs $+$ and $-$ refer to neutrons and protons respectively.

The second set \cite{chp} has somewhat different parameters, but gives
as good fit to the single-particle level positions. It is:
\begin{eqnarray} \nonumber \label{chpls}
R = R_{ls} = 1.24A^{1/3}\; fm,\;\;\; a = a_{ls} = 0.63\; fm, \\
U_0 = (-53.3 \pm 33.6\,\frac{N-Z}{A})\; MeV,\;\;\; U_{ls} = -0.263\,(1+2\,
\frac{N-Z}{A})\,U_0
\end{eqnarray}

In both cases we assume the same values of the density parameters \cite{bm}:
\begin{equation}
R = 1.11\,A^{1/3}\;fm;\;\;\; a = 0.54\; fm\;\;\; \rho_0 = 0.17\; fm^{-3}.
\end{equation}

The results of calculations obtained for both sets are listed in Table
\ref{t:sc}.  The main contribution to AM, as expected, comes from the spin
current. The variation of the parameters of potential changes this term by
3 -- 6\%.
The next in magnitude term is due to the potential part of
spin-orbit current, it constitutes about 30\% of the spin current contribution.

In the calculations based on the sets of parameters (\ref{bmls})
and (\ref{chpls}) the LS-potential was not consistent with the
two-particle LS-interaction (\ref{eq:so}). So, the potential LS-contribution
to the AM was calculated in two ways. First, starting from the LS-potential
itself.
Second, it was obtained from the spin-orbit potential constructed from the
two-particle interaction (\ref{eq:so}) - (\ref{uls2}).  The difference
between second and first ways of calculation is listed as $\delta
\kappa^0_{ls}$ in Table \ref{t:sc}. It is only natural to perform
a consistent calculation, choosing the amplitude of LS-potential in
accordance with (\ref{uls}), (\ref{uls2}), and its radius and
diffuseness parameters the same as those of nuclear density. In this way we
come
to the third, LS-consistent, version of the potential:
\begin{eqnarray} \nonumber \label{csls}
R = 1.27A^{1/3}\; fm,\;\;\; a = 0.67\; fm, \;\;\;
U_0 = (-51 \pm 33\,\frac{N-Z}{A})\; MeV, \\
R_{ls} = 1.11A^{1/3}\; fm,\;\;\; a_{ls} = 0.54\; fm, \;\;\;
U_{ls} = (14.5 \mp 8.3\,\frac{N-Z}{A})\; MeV.
\end{eqnarray}
We kept the central potential parameters as in (\ref{bmls}) because it gave
smaller $\delta \kappa_{ls}^0$.
The calculation based on this version of the potential give
$\delta \kappa_{ls}^0 = 0$.

It is noteworthy that the sum of all contributions to AM is less
sensitive to a specific potential than each of them taken separately. Its
variation
does not exceed 5\% when going from one set of parameters to another.

We believe that the results obtained with the LS-consistent potential are
the most reliable ones. They are presented in the last
line of Table \ref{tab:k}.

\section{Discussion of the results}
Let us discuss now why the present results differ from those obtained in
Refs.\cite{fks,bp,bp1}.

As distinct from Ref. \cite{fks}, we have included now the contribution of the
spin-orbit current which turns out quite essential.

On the other hand, the Woods-Saxon potential used here for numerical
calculations is more realistic than the oscillator one used for those
calculations in Refs. \cite{bp,bp1}. Then, in
Refs. \cite{bp,bp1} the spin-orbit interaction is treated perturbatively,
while in the present work it is treated exactly. Let us note here again that
the oscillator potential without spin-orbit interaction allows one to get an
exact analytic solution for the AM, as it has been done in Section 3 of the
present work.

To summarize the comparison we wish to say that the above arguments can serve
as serious grounds to consider the numerical results of the present work to
be the most reliable ones for the single-particle approximation.

However, accurate quantitative predictions for nuclear AM's cannot be made
without proper treatment of nuclear many-body effects. As to the core
excitations by the weak interaction Hamiltonian (\ref{2b}), their
contribution to nuclear AM has been demonstrated to be small \cite{fks1}.
However, the problem of the configurations mixing caused by usual P-even
residual nucleon-nucleon interaction is here much more serious than in the
case of nuclear magnetic moments (or the second neutral current constant
\cite{bp}). The point is that in the last cases it is the specific,
kinematical, nature of the operator $\vec{\sigma}$ which allows one to
restrict with a reasonable accuracy to the excitations within a spin-orbit
doublet only. As to the AM, even if the multiconfiguration problem could be
reduced to the calculations with the effective operator (\ref{eq:lo})
(which demands a special proof by itself), that operator is
coordinate-dependent and in this sense resembles more the magnetic
octupole operator than the magnetic dipole one. And many-body effects
can renormalize M3 operator much more strongly than M1 operator. A drastic
example in this
respect is $\;^{209}$Bi where the one-particle value of the magnetic
octupole moment is almost four times smaller than the experimental one
\cite{dt}.

Let us note here that due to the theoretical uncertainties discussed, in
the situation when the AM contribution dominates essentially atomic
nuclear-spin-dependent PNC effects, the proposal \cite{bp} to single out
in those effects the true neutral current contribution by combining
experimental data from various heavy nuclei, does not look realistic.

In conclusion it should be emphasized however that even with all those
theoretical uncertainties kept in mind, the problem of experimental
observation of a new physical phenomenon, nuclear anapole moment, is a
fascinating one. Moreover, if the theoretical results of one-particle
approximation, which are by themselves remarkably stable by nuclear
standards, will be supplemented by a serious treatment of many-body
effects, those experimental investigations will give reliable quantitative
information on P-odd nuclear forces.

{\bf Acknowledgements}

One of the authors (I.B.Kh.) wishes to thank the Institute for Nuclear Theory,
University of Washington, and the theoretical physics groups of the
Universities
of New South Wales, Melbourne and Cambridge,
where part of this work was done, for their kind hospitality.

\pagebreak

\newpage


 \begin{table}[b]
  \begin{center}
  \begin{tabular}{lcccc}
Nucleus         &$^{133}$Cs&$^{203,205}$Tl  &$^{209}$Bi  &$^{207}$Pb\\
& & & & \\
Analytical result \protect\cite{fks}&$0.08g_p$
&$0.11g_p$&$0.11g_p$&$-0.08g_n$\\
Its value at $g_p=4$ \protect\cite{fks}&$0.33$    &$0.44$  &$0.45$    &\\
Numerical result \protect\cite{fks} &$0.06g_p$  &$0.09g_p$
&$0.08g_p$&$-0.09g_n$\\
Its value at $g_p=4$ \protect\cite{fks}&$0.25$      &$0.38$&$0.31$    &\\
\protect\cite{hhm}                    &$0.14(0.28)$& -    & -        & - \\
\protect\cite{bp,bp1}                 &$0.24$      &$(0.24)$&$(0.24)$& - \\
This work, analytical, $g_p=4.5$  &$0.33$ &$0.42$ &$0.45$ &$-0.09g_n+0.005$\\
This work, numerical, $g_p=4.5$   &$0.26$   &$0.40$&$0.29$&$-0.10g_n+0.004$\\
        \end{tabular}
 \end{center}
  \caption{Effective constants $\kappa$ for heavy nuclei}
  \label{tab:k}
  \end{table}

  \begin{table}[b]
  \begin{center}
  \begin{tabular}{ccccccc}
 Nucleus &$\langle r^2_m\rangle _{exp}$&$\frac{3}{5}r^2_0 A^{2/3}$&$\rho^2
(N+3/2)$\\
 & & & \\
 $^{41}$Ca$_{20}$ & $15.92(0.48)$ \protect\cite{pl}& $10.27$ & $15.61$ \\
 $^{45}$Sc$_{21}$ & $12.67(1.57)$ \protect\cite{dv}& $10.93$ & $16.11$ \\
 $^{51}$V$_{23}$  & $12.89(0.72)$ \protect\cite{dv}& $11.88$ & $16.79$ \\
 $^{59}$Co$_{27}$ & $13.99(1.05)$ \protect\cite{dv}& $13.09$ & $17.62$ \\
        \end{tabular}
  \end{center}
  \caption{Mean squared magnetic radii}
  \label{tab:r}
  \end{table}
 \begin{table}[b]
  \begin{center}
  \begin{tabular}{lcccc}
Nucleus         &$^{133}$Cs&$^{203,205}$Tl  &$^{209}$Bi  &$^{207}$Pb\\
& & & & \\
$\kappa_s\;\;^{1)}$& 0.317&0.485&0.376&-0.099$g_n$\\
$\kappa_s\;\;^{2)}$& 0.301&0.463&0.349&-0.095$g_n$\\
$\kappa_s\;\;^{3)}$& 0.310&0.497&0.353&-0.099$g_n$\\
$\kappa^0_{ls}\;\;^{3)}$&-0.088&-0.120&-0.126&0\\
$\kappa^1_{ls}\;\;^{3)}$&0.014&0.018&0.019&-0.006$g_n$\\
$\delta \kappa^0_{ls}\;\;^{1)}$&0.003&0.019&-0.003&0\\
$\delta \kappa^0_{ls}\;\;^{2)}$&0.01&0.032&0.014&0\\
$\kappa_{conv}\;^{3)}$&-0.019&-0.055&-0.019&0\\
$\kappa^0_c\;\;^{3)}$&0.048&0.064&0.070&0\\
$\kappa^1_c\;\;^{3)}$&-0.007&-0.008&-0.009&0.004\\
$\kappa\;\;^{1)}$& 0.268&0.396&0.309&-0.105$g_n$ + 0.004\\
$\kappa\;\;^{2)}$& 0.258&0.383&0.298&-0.101$g_n$ + 0.004\\
$\kappa\;\;^{3)}$& 0.257&0.396&0.289&-0.105$g_n$ + 0.004\\
\end{tabular}
\end{center}
\caption{Different contributions to the effective constant $\kappa $.
\protect\newline $^{1)}\;\;$  The potential parameters from
Ref. \protect\cite{bm} (see eq. (\protect\ref{bmls})). \protect\newline
$^{2)}\;\;$ The potential parameters from
Ref. \protect\cite{chp} (see eq. (\protect\ref{chpls})). \protect\newline
$^{3)}\;\;$ The consistent LS-parameters  (see eq. (\protect\ref{csls})).}

\label{t:sc}
\end{table}

\newpage

\begin{figcap}



\item{  First order correction $\delta R(r)$ to
radial wave function
of $1h_{9/2}$ level. Dotted line corresponds to constant
density and no spin-orbit potential ($\delta R = rR(r)$). Dash-dotted line
refers to varying
density and no spin-orbit potential. Dashed line refers to constant density,
spin-orbit potential included. Full line corresponds to full potential and
varying density.}
\end{figcap}


\begin{thebibliography}{99}

\bibitem{khr} I.B. Khriplovich: Parity Nonconservation in Atomic Phenomena
(Gordon and Breach, London, 1991)
\bibitem{fk} V.V. Flambaum, I.B. Khriplovich: Zh.Eksp.Teor.Fiz. {\bf 79}
(1980) 1656 [Sov.Phys. JETP {\bf 52} (1980) 835]
\bibitem{fks} V.V. Flambaum, I.B. Khriplovich, O.P. Sushkov: Phys.Lett.
{\bf B145} (1984) 367
\bibitem{zel} Ya.B. Zel'dovich: Zh.Eksp.Teor.Fiz. {\bf 33}
(1957) 1531 [Sov.Phys. JETP {\bf 6} (1957) 1184] (This paper contains also
the mention of the analogous results obtained by V.G. Vaks.)
\bibitem{kp} I.B. Khriplovich, M.E. Pospelov: Z.Phys. {\bf D17}
(1990) 81
\bibitem{hhm} W.C. Haxton, E.M. Henley, M.J. Musolf: Phys.Rev.Lett.
{\bf 63} (1989) 949
\bibitem{bp} C. Bouchiat, C.A. Piketty: Z.Phys. {\bf C49} (1991) 91
\bibitem{bp1} C. Bouchiat, C.A. Piketty: Phys.Lett. {\bf B269} (1991) 195;
erratum {\bf B274} (1992) 526
\bibitem{wie} M.S. Noecker, B.P. Masterson, C.E. Wieman: Phys.Rev.Lett.
{\bf 61} (1988) 310
\bibitem{cur} F. Curtis Michel: Phys.Rev. {\bf 133B} (1964) 329
\bibitem{dep} B. Desplanque: Phys.Lett. {\bf B47} (1973) 212
\bibitem{fks1} V.V. Flambaum, I.B. Khriplovich, O.P. Sushkov: Phys.Lett.
{\bf B162} (1985) 213; Nucl.Phys. {\bf A449} (1986) 750
\bibitem{ddh} B. Desplanque, J.F. Donoghue, B.R. Holstein: Ann.Phys.
{\bf 124} (1980) 449
\bibitem{dfst} V.F. Dmitriev, V.V. Flambaum, O.P. Sushkov, V.B. Telitsin:
Phys.Lett. {\bf B125} (1983) 1
\bibitem{fts} V.V. Flambaum, V.B. Telitsin, O.P. Sushkov: Nucl.Phys.
{\bf A444} (1985) 611
\bibitem{st} O.P. Sushkov, V.B. Telitsin: Phys.Rev. {\bf C48} (1993)
\bibitem{mck} B.H.J. McKellar, Phys.Rev.Lett. {\bf 20} (1968) 1542
\bibitem{lob} G.A. Lobov: Izv.Akad.Nauk SSSR (Ser.Fiz.) {\bf 44} (1980) 2364
\bibitem{ah} E.G. Adelberger, W.C. Haxton: Ann.Rev.Nucl.Part.Sci. {\bf 35}
(1985) 501
\bibitem{bm} A. Bohr and B.R. Mottelson: Nuclear Structure, v.1 (W.A.
Benjamin, Inc., New York, Amsterdam, 1969)
\bibitem{pag} V.I. Pagurova: Tables of Incomplete Gamma-Function (Moscow, 1963)
\bibitem{pl} S. Platchkov et al: Phys.Rev.Lett. {\bf 61} (1988) 1465
\bibitem{dv} H. De Vries, G.J.G. Van Niftric, L. Lapikas, Phys.Lett.
{\bf 33B} (1970) 403
\bibitem{dt} V.F. Dmitriev, V.B. Telitsin: Nucl.Phys. {\bf A402} (1983) 581
\bibitem{chp} V.A. Chepurnov: Yad.Fiz. {\bf 6} (1967) 955 [Sov. J. Nucl. Phys.
{\bf 6} (1967)]
\bibitem{sad74} B.I. Birbrair and V.A. Sadovnikova: Yad.Fiz. {\bf 20} (1974)
347 [Sov. J. Nucl. Phys. {\bf 20} (1974)]
\end{thebibliography}
\end{document}